\documentclass[10pt,superscriptaddress,tightenlines,twocolumn,amsmath,amssymb,aps, pra]{revtex4-2}
\usepackage{amsfonts}
\usepackage{amsmath}
\usepackage{amsthm}
\usepackage{amscd}
\usepackage{amssymb}
\usepackage{subfigure}
\usepackage{amsxtra}
\usepackage{gensymb}
\usepackage{bm}           
\usepackage{bbm}
\usepackage{graphicx}
\usepackage{epstopdf}
\usepackage{color}
\usepackage{hyperref}
\usepackage{makecell}
\usepackage{multirow}
\usepackage{lettrine}

\hypersetup{colorlinks=true, linkcolor=blue, citecolor=red, urlcolor=blue}

\def\bi#1\ei {\begin{itemize}#1\end{itemize}}
\def\bn#1\en {\begin{enumerate}#1\end{enumerate}}
\def\bea#1\eea {\begin{align}#1\end{align}}
\def\bean#1\eean {\begin{align*}#1\end{align*}}
\def\ben#1\een {\begin{equation*}#1\end{equation*}}
\def\be#1\ee {\begin{equation}#1\end{equation}}
\def\bes#1\ees {\begin{equation}\begin{split}#1\end{split}\end{equation}}
\def\bear#1\eear {\begin{eqnarray}#1\end{eqnarray}}
\def\bear#1\eear {\begin{eqnarray*}#1\end{eqnarray*}}

\newcommand{\beq}{\begin{equation}}
\newcommand{\eeq}{\end{equation}}

\begin{document}

\title{Quantum Neural Network for Quantum Neural Computing}
\author{Min-Gang Zhou}\thanks{These authors contributed equally to this work}
\affiliation{National Laboratory of Solid State Microstructures, School of Physics, Collaborative Innovation Center of Advanced Microstructures, Nanjing University, Nanjing 210093, China}
\author{Zhi-Ping Liu}\thanks{These authors contributed equally to this work}
\affiliation{National Laboratory of Solid State Microstructures, School of Physics, Collaborative Innovation Center of Advanced Microstructures, Nanjing University, Nanjing 210093, China}
\author{Hua-Lei Yin}\thanks{These authors contributed equally to this work}
\affiliation{National Laboratory of Solid State Microstructures, School of Physics, Collaborative Innovation Center of Advanced Microstructures, Nanjing University, Nanjing 210093, China}
\author{Chen-Long Li}
\affiliation{National Laboratory of Solid State Microstructures, School of Physics, Collaborative Innovation Center of Advanced Microstructures, Nanjing University, Nanjing 210093, China}
\author{Tong-Kai Xu}
\affiliation{MatricTime Digital Technology Co. Ltd., Nanjing 211899, China}
\author{Zeng-Bing Chen}
\email{zbchen@nju.edu.cn}
\affiliation{National Laboratory of Solid State Microstructures, School of Physics, Collaborative Innovation Center of Advanced Microstructures, Nanjing University, Nanjing 210093, China}
\affiliation{MatricTime Digital Technology Co. Ltd., Nanjing 211899, China}

\date{\today}

\begin{abstract}
Neural networks have achieved impressive breakthroughs in both industry and academia. How to effectively develop neural networks on quantum computing devices is a challenging open problem. Here, we propose a new quantum neural network model for quantum neural computing using (classically-controlled) single-qubit operations and measurements on real-world quantum systems with naturally occurring environment-induced decoherence, which greatly reduces the difficulties of physical implementations. Our model circumvents the problem that the state-space size grows exponentially with the number of neurons, thereby greatly reducing memory requirements and allowing for fast optimization with traditional optimization algorithms. We benchmark our model for handwritten digit recognition and other nonlinear classification tasks. The results show that our model has an amazing nonlinear classification ability and robustness to noise. Furthermore, our model allows quantum computing to be applied in a wider context and inspires the earlier development of a quantum neural computer than standard quantum computers.
\end{abstract}

\maketitle

\bigskip
\noindent
\textbf{Introduction}

Developing new computing paradigms \cite{zadeh1996fuzzy,amit2018artificial,nielsen2002quantum,zhang2020system} has attracted considerable attention in recent years due to the increasing cost of computing and the von Neumann bottleneck \cite{waldrop2016chips}. Conventional (hard) computing is characterized by precision, certainty, and
rigor. In contrast, ``soft computing'' \cite{zadeh1996fuzzy,amit2018artificial} is a newer approach to computing that mimics human
thinking to learn and reason in an environment of imprecision, uncertainty,
and partial truth. This approach aims to address real-world complexities with tractability, robustness, and low solution
costs. In particular, neural networks (NNs), a subfield of soft computing, have rapidly evolved in both theory and practice during the current machine learning boom \cite{goodfellow2016deep,nielsen2015neural}. With backpropagation algorithms, NNs have achieved impressive breakthroughs in both industry and academia \cite{jordan2015machine,bishop2006pattern} and may even alter the way computation is performed \cite{zhang2020system}. However, the training cost of NNs can become very expensive as the network size increases \cite{brown2020language}. More seriously, it is difficult for NNs to simulate quantum many-body systems with exponentially large quantum state spaces \cite{biamonte2017quantum}, which restricts basic scientific research and the intelligent development of biopharmaceutical and material design.

Quantum computing \cite{nielsen2002quantum} is another paradigm shift in
computing, and it promises to solve the aforementioned difficulties of NNs. How to effectively develop NNs on quantum computing devices is a challenging open problem \cite{biamonte2017quantum,schuld2014quest,zhou2022experimental} that is still in its initial stages of exploration. In recent years, many novel and original works have attempted to develop well-performing quantum NN models \cite{wan2017quantum,beer2020training,bondarenko2020quantum,cong2019quantum,jiang2021co,mcclean2018barren,farhi2018classification,sharma2022trainability,da2016quantum,torrontegui2019unitary,herrmann2022realizing,huang2022quantum} on noisy intermediate-scale quantum devices \cite{preskill2018quantum}, and these networks can be used to learn tasks involving quantum data or to improve classical models. However, despite the remarkable progress in the physical implementation of quantum computing in recent years, a number of significant challenges remain for building a large-scale
quantum computer \cite{kjaergaard2020superconducting,ladd2010quantum,barends2014superconducting}. Thus, if the quest for quantum NNs heavily relies on standard quantum computing devices, the scope of applying quantum NNs might be quite restrictive.

A real-world quantum system is always characterized by nonunitary, faulty evolutions and is coupled with a noisy and dissipative environment. The real-system complexities in the quantum domain call for a new paradigm of quantum computing aiming at nonclassical computation using real-world quantum systems. Thus, the new quantum computing paradigm, called soft quantum computing to be compared with classical soft computing, deals with classically intractable computation under the conditions of noisy and faulty quantum evolutions and measurements, while being tolerant of those effects that are detrimental for the standard quantum computing paradigm.

Here, we propose for the first time a quantum NN model to illustrate soft quantum computing. Unlike other quantum NN models, we develop NNs for quantum neural computing based on ``soft quantum neurons'', which are building blocks of soft quantum computing and subject to only single-qubit operations, classically-controlled single-qubit operations and measurements, thus significantly reducing the difficulties of physical implementations. We demonstrate that quantum correlations characterized by non-zero quantum discord are present for quantum neurons in our model. The simulation results show that our quantum perceptron can be used to classify nonlinear problems and simulate the XOR gate. In contrast, classical perceptrons do not possess such nonlinear classification capabilities. Furthermore, our model is able to classify handwritten digits with an extraordinary generalization ability even without hidden layers. Our model also has a significant accuracy advantage over other quantum NNs for the abovementioned tasks. Prominently, the proposed soft quantum neurons can be integrated into quantum analogues of typical topological architectures \cite{krizhevsky2012imagenet,scarselli2008graph,goodfellow2014generative,hochreiter1997long} in classical NNs. The respective advantages of quantum technology and classical network architectures can thus be well combined in our quantum NN model.

\bigskip
\noindent
\textbf{Results}

\noindent
\textbf{Soft quantum neurons.} 

\begin{figure}
\includegraphics[width=7cm]{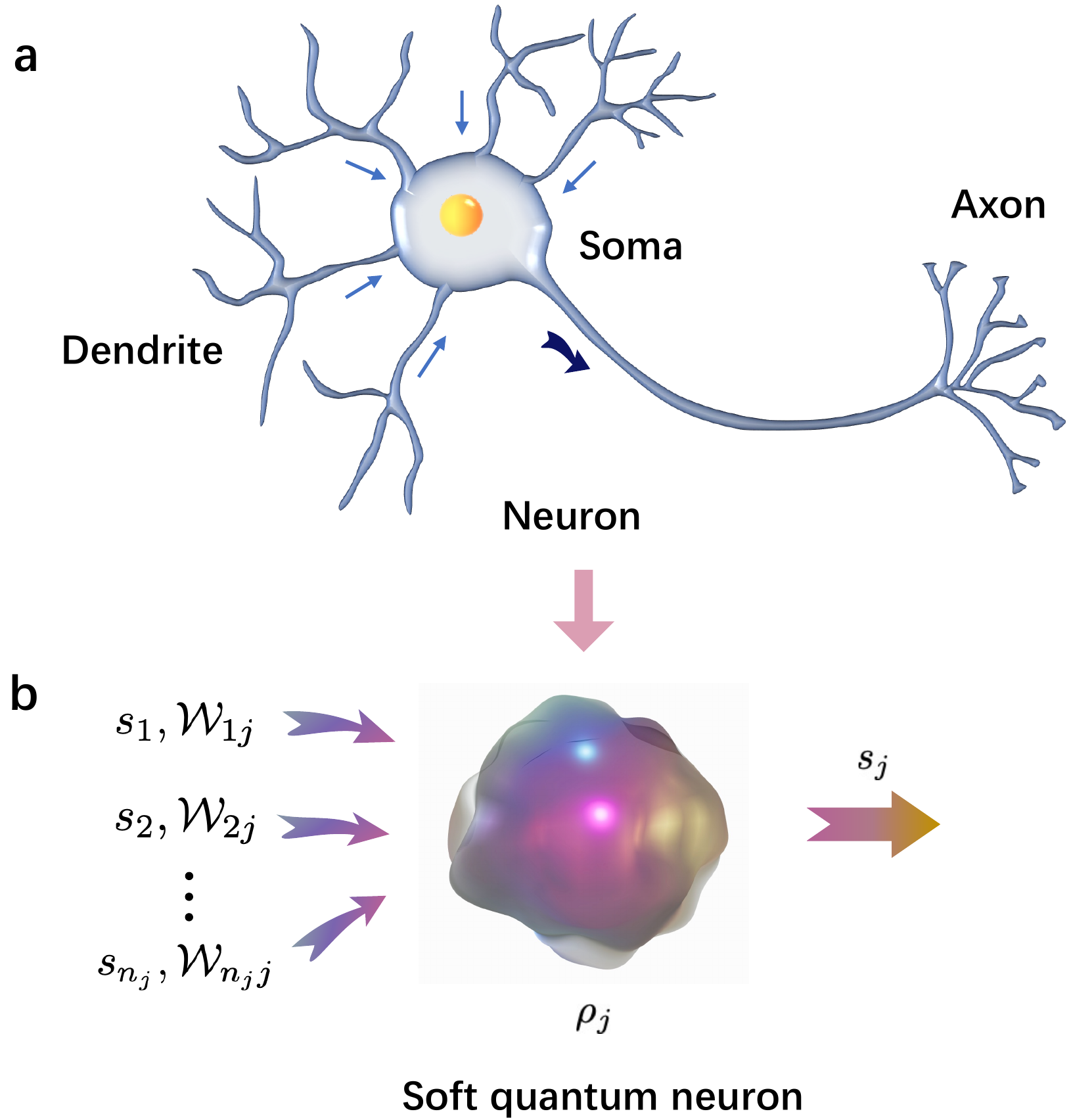}
\caption{\textbf{A drastically simplified drawing of a neuron and the soft quantum neuron model. a} The neuron integrates hundreds or thousands of impinging signals through its dendrites. After processing by the cell body, the neuron outputs a signal through its axon to another neuron for processing in the form of an action potential when its internal potential exceeds a certain threshold. \textbf{b} Similarly, a ``soft quantum neuron'' can in principle receive hundreds or thousands of input signals. These signals affect the evolution of the soft quantum neuron. The evolved soft quantum neuron is measured and decides whether to output a signal according to the measurement result.}
\label{qneuron}
\end{figure}

Quantizing the smallest building block of classical NNs, namely, the neuron, is a key challenge in building quantum NNs. Our soft quantum neuron model is inspired by biological neurons (Fig.~\ref{qneuron}), which can be implemented on realistic quantum systems. The term ``soft" utilized here highlights the ability of our model to handle realistic environments and evolutions, distinguishing it from the standard quantum computing models. It is worth noting that soft computing is a proprietary term that is conceptually opposite to hard computing. In our proposal, a quantum neuron is modelled by a noisy qubit, which can be coupled with its surrounding environment. The initial state of the $j$th neuron can be described by a density matrix $\rho_j^{in}$ in the computational basis $\left\vert 0\right\rangle $ and $\left\vert 1\right\rangle $. The quantum neuron $\rho_j^{in}$ accepts $n_{j}$ outputs $s_{i}$ $(i=1,2,...,n_{j}$) from the final states $\rho_i^{out}$ $(i=1,2,...,n_{j}$) of the other possible $n_{j}$ neurons. The output $s_i$ is determined by a $2-$outcome projective measurement on $\rho_{i}^{out}$ in the computational basis. It is therefore a classical binary signal, namely, $s_{i}=0$ or $1$. When $s_{i}=1$, corresponding to the case where $\rho_i^{out}$ is measured and collapses to the state $|1\rangle$, the quantum neuron $\rho_{j}^{in}$ is acted upon by an arbitrary superoperator $\mathcal{W}_{ij}$, while when $s_{i}=0$ ($\rho_i^{out}$ collapses to the state $|0\rangle$), nothing happens to $\rho_{j}^{in}$. Ideally, $\mathcal{W}_{ij}$ can be replaced by a corresponding unitary operator $W_{ij}$. As a result, the evolution of the whole system from the state $\bigotimes\nolimits_{i=1}^{n_{j}}\rho_{i}^{out}\otimes\rho
_{j}^{in}$ is
\begin{align}
\rho^{mid}_{\left\{  i\right\}  j}  &  =\mathcal{T}\bigotimes\nolimits_{i=1}%
^{n_{j}}\mathcal{O}_{ij}(\rho_{i}^{out}\otimes\rho_{j}^{in})\nonumber\\
&  \equiv\mathcal{T}\bigotimes\nolimits_{i=1}^{n_{j}}[\mathcal{P}_{\left\vert
0\right\rangle _{i}}\otimes\hat{I}_{j}+\mathcal{P}_{\left\vert 1\right\rangle
_{i}}\otimes\mathcal{W}_{ij}](\rho_{i}^{out}\otimes\rho_{j}^{in}).
\label{evolu}%
\end{align} 
Here, $\mathcal{O}_{ij}$ is a classically-controlled single-qubit operation, the superprojectors $ \mathcal{P}%
_{\left\vert s\right\rangle }$ are defined by $\mathcal{P}_{\left\vert
s\right\rangle }\rho=\left\vert s\right\rangle \left\langle s\right\vert
\rho\left\vert s\right\rangle \left\langle s\right\vert $, $\hat{I}$ is the
identity operator, and $\mathcal{T}$ represents a time-ordering operation. All $\mathcal{W}_{ij}$ act upon the target neuron $\rho_j^{in}$ with
specific temporal patterns. As different quantum operations $\mathcal{W}_{ij}$ might be noncommutative, the time-ordering of these operations is important. The state of the target neuron after the evolution of Eq.~(\ref{evolu}) can be obtained by tracing out all the input neurons $\rho_i^{out}$, namely, $\rho^{mid}_{j}=\mathrm{tr}_{\left\{  i\right\}  }\rho^{mid}_{\left\{  i\right\}
j}=\mathcal{T}\prod_{i=1}^{n_{j}}[p_{i}\hat{I}_{j}+(1-p_{i}%
)\mathcal{W}_{ij}]\rho_{j}^{in}$, where $p_{i}\equiv p_{i}(0)=\mathrm{tr}%
(\left\vert 0\right\rangle _{i}\left\langle 0\right\vert \rho_{i}^{out})$.

After the evolution of Eq.~(\ref{evolu}), the target neuron $\rho^{mid}_{j}$ is independently acted upon by a local bias superoperator $\mathcal{U}_{j}$. This operator is designed to improve the flexibility and learning ability of quantum neurons. Ideally, $\mathcal{U}_{j}$ can be replaced by a corresponding unitary operator $U_{j}$. The action of $\mathcal{U}_{j}$ is similar to adding bias to neurons in classical NNs \cite{goodfellow2016deep,nielsen2015neural}. The final state of the target neuron is thus $\rho_j^{out} = \mathcal{U}_j(\rho^{mid}_j)$. Similarly, the output $s_j$ of the target neuron is obtained by a $2-$outcome projective measurement on $\rho_{j}^{out}$ in the computational basis. The output signal $s_{j}$ of the target neuron is
\begin{equation}
s_{j}=\left\{
\begin{array}
[c]{ll}%
0\text{ \ \ } & \text{with probability }p_{j}(0)\\
1\text{ \ \ } & \text{with probability }1-p_{j}(0).
\end{array}
\right.  \label{sj}%
\end{equation} 
The output $s_j$ can be accepted by all other connecting quantum neurons and affects the evolution of the quantum neurons that accept the output. This completes the specification of our proposed quantum neuron model. Strikingly, our model contains noisy cases, which allow our model to work under the conditions of noisy and faulty quantum evolutions and measurements. An elementary setup of our model is the soft quantum perceptron,
which consists of a soft quantum neuron accepting inputs of $n$ other soft quantum neurons and providing a single output, though in probability.\\

\bigskip
\noindent
\textbf{Quantumness of quantum neurons.} 

All final states of our quantum neurons are mixed states, as the evolution of these neurons depends on the measurements of their input neurons, thus introducing classical probability. Although such measurements make the neurons evolve into mixed states, the proposed quantum 
neurons can still develop quantum correlations arising from quantum discord. To make this clear, we consider the simplest two-neuron case. For the two neurons in the states\ $\rho_{1}^{out}=p_{1}%
\left\vert 0\right\rangle _{1}\left\langle 0\right\vert +(1-p_{1})\left\vert
1\right\rangle _{1}\left\langle 1\right\vert $\ ($p_{1}\neq0,1$) and $\rho
_{2}^{in}$, the action of an operation $\mathcal{O}_{12}$
results in the state
\begin{align}
\rho^{mid}_{12}  &  =(\mathcal{P}_{\left\vert 0\right\rangle _{1}}\otimes
\hat{I}_{2}+\mathcal{P}_{\left\vert 1\right\rangle _{1}}\otimes \mathcal{W}%
_{12})(\rho_{1}^{out}\otimes\rho_{2}^{in})\nonumber\\
&  =p_{1}\left\vert 0\right\rangle _{1}\left\langle 0\right\vert \otimes
\rho_{2}^{in}+(1-p_{1})\left\vert 1\right\rangle _{1}\left\langle 1\right\vert
\otimes \mathcal{W}_{12}(\rho_{2}^{in}), \label{tn}%
\end{align}
where $\mathcal{W}_{12}$ represents a specific quantum channel. Quantum correlations, if
any, of $\rho^{mid}_{12}$ can be quantified by the quantum discord
\cite{ollivier2001quantum}. Any bipartite state is called fully classically correlated if
it is of the form \cite{streltsov2011behavior} 
$\rho_{12}^{c}=\sum_{i,j}p_{ij}\left\vert i\right\rangle _{1}\left\langle
i\right\vert \otimes\left\vert j\right\rangle _{2}\left\langle j\right\vert$;
otherwise, it is quantum correlated. Here, $\left\vert i\right\rangle _{1}$ and
$\left\vert j\right\rangle _{2}$ are the orthonourmal bases of the two parties,
with nonnegative probabilities $p_{ij}$.

Obviously, for $\rho^{mid}_{12}$ in Eq.~(\ref{tn}) the
first neuron becomes quantum-correlated with the second
as long as $\mathcal{W}ƒ_{12}(\rho_{2}^{in})$ and $\rho_{2}^{in}$ are
nonorthogonal \cite{dakic2010necessary,ciccarello2012creating,lanyon2013experimental}. In particular, Refs.~\cite{ciccarello2012creating,lanyon2013experimental}
show the creation of discord, from classically correlated two-qubit states, by
applying an amplitude-damping process only on one of the qubits; for the
phase-damping process, see Ref.~\cite{streltsov2011behavior}. Actually, $\rho^{mid}_{12}$
in Eq.~(\ref{tn}) is the \textit{classical-quantum state}, as dubbed in
Ref.~\cite{ciccarello2012creating}. While for measurements on neuron-1 the discord is
zero, measurements on neuron-2 in general lead to nonzero discord.

\begin{figure}
	\centering
	\includegraphics[width=8cm]{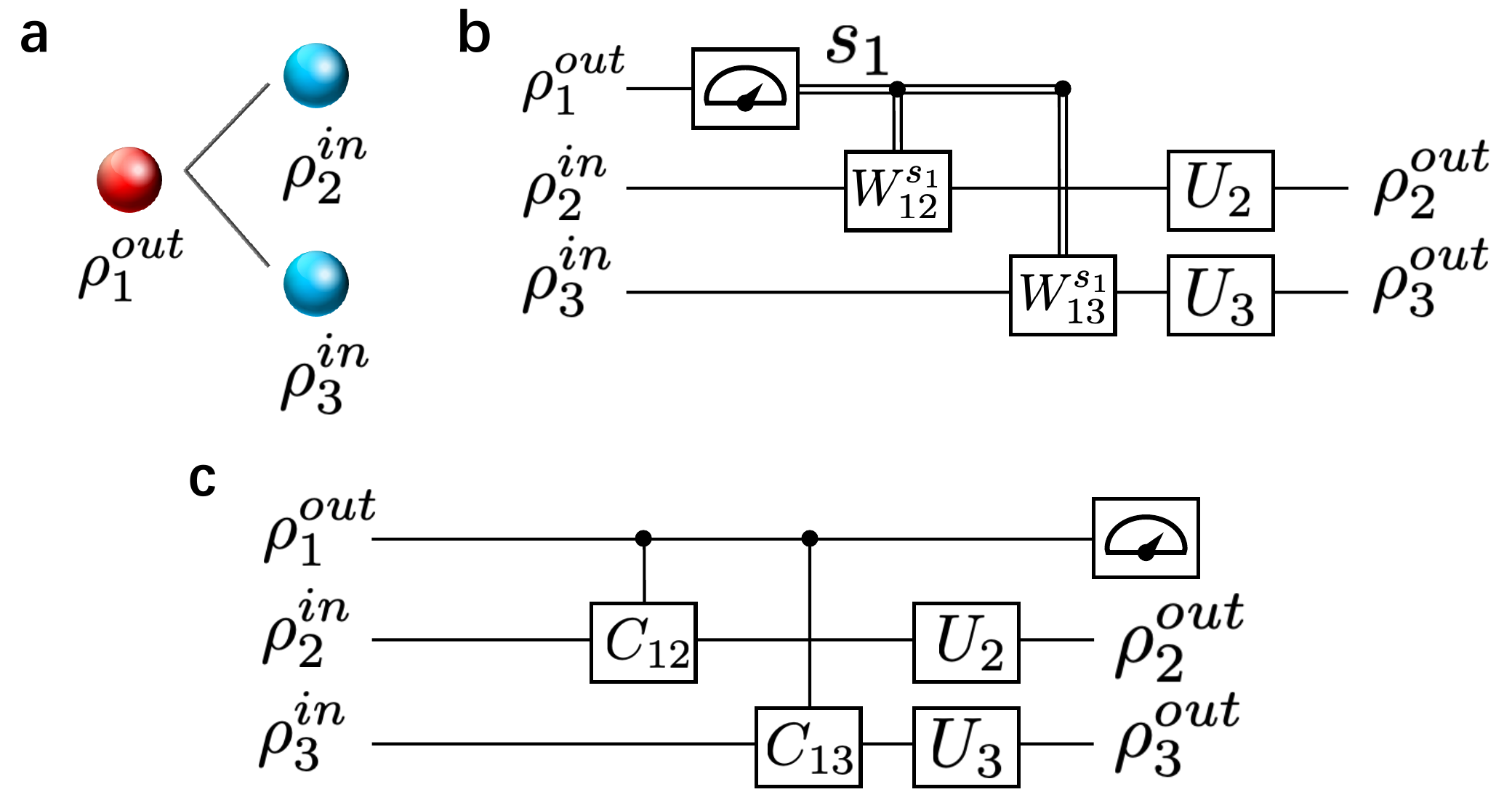}
	\caption{\textbf{Simplified demonstration of the principle of deferred measurement as applied to} our model. a} A simple example of our model. The quantum neuron $\rho_1^{out}$ sends signals to $\rho_2^{in}$ and $\rho_3^{in}$. \textbf{b} The quantum circuit model of (\textbf{a}). \textbf{c} The deferred measurement quantum circuit of (\textbf{a}), which is equivalent to (\textbf{b}). Here, $C_{12}$ and $C_{13}$ are controlled unitary operators acting on $\rho_2^{in}$ and $\rho_3^{in}$, respectively.
	\label{deferred}
\end{figure}

Thus, we reveal a crucial property of our quantum neuron model. Namely, quantum correlations arising from quantum discord can be developed between the proposed quantum neurons in our model, although these neurons are generally in mixed states. Note that the existing quantum neural network models are mainly based on variational quantum circuits requiring two-qubit gates.

More remarkably, our model can be equated to quantum circuits generating quantum entanglement. To illustrate this more clearly, we take three neurons in Fig.~\ref{deferred}\textbf{a} as an example and represent their interactions by a quantum circuit model shown in Fig.~\ref{deferred}\textbf{b}. To facilitate the demonstration, we consider the ideal case where $\rho_2^{in}$ and $\rho_3^{in}$ are pure states and $\mathcal{W}_{ij}$ and $\mathcal{U}_j$ are replaced by corresponding unitary operators $W_{ij}$ and $U_{j}$, respectively. According to the principle of deferred measurement \cite{nielsen2002quantum}, measurements can always be moved from the middle step of a quantum circuit to the end of the circuit. Therefore, the circuit in Fig.~\ref{deferred}\textbf{c} is equivalent to that in Fig.~\ref{deferred}\textbf{b}. In the equivalent circuit, the unitaries that are conditional on the measurement results are replaced by controlled unitary operations on $\rho_2^{in}$ and $\rho_3^{in}$. It is easy to verify that quantum entanglement can exist between neuron-1 and neuron-2 (as well as neuron-3) in Fig.~\ref{deferred}\textbf{c}. Another example of the principle of deferred measurement can be found in teleportation \cite{nielsen2002quantum}. Nonetheless, it remains unclear whether this equivalence can be effectively utilized in computing tasks. We leave this matter for future work.\\

\bigskip
\noindent
\textbf{Soft quantum neural network.} 

Quantum neurons are connected together in various configurations to form quantum NNs with learning abilities, thus representing a quantum neural computing device obeying the evolution-measurement rules provided above. Our neurons can in principle be combined into quantum analogues of any classical network architecture that has proven effective in many applications. In this work, we present a fully-connected soft quantum feedforward NN (SQFNN) for application to supervised learning.

Neurons are arranged in layers in a fully-connected feedforward NN (FNN). Each neuron accepts all the signals sent by the neurons in the previous layer and outputs the integrated signal to each neuron in the next layer. Note that there is no signal transmission between neurons within the same layer. To date, there has been no satisfactory quantum version of this simple model. Because no neuron can perfectly copy its quantum state in multiple duplicates as an output to the next layer due to the quantum 
no-cloning theorem \cite{wootters1982single}, the output is not perfectly shared by neurons in the next layer. Because of the same theorem, quantum neural computing and standard quantum computing have incompatible requirements that are difficult to reconcile \cite{schuld2014quest}. Our quantum NN model resolves this incompatibility by measuring each soft quantum neuron to give classical information as the integrated signal. This feature is essential for our model to be a genuine quantum NN model, which, while incorporating a neural computing mechanism, uses quantum laws consistently throughout neural computing. 

In fact, many studies have made bold attempts in this challenging area. For example, Ref.~\cite{wan2017quantum} introduces a general ``fan-out" unit that distributes information about the input state into several output qubits. The quantum neuron in Ref. \cite{beer2020training} is modelled as an arbitrary unitary operator with $m$ input qubits and $n$ output qubits. These attempts provide new perspectives for resolving the abovementioned incompatibility. Unfortunately, none of them directly confronts this incompatibility. The neurons in these schemes still cannot share the outputs of the neurons in the previous layer; conversely, each neuron can only send different signals to different neurons in the next layer. In that sense, our NN is quite different from these quantum NNs.

\begin{figure}
	\includegraphics[width=6.5cm]{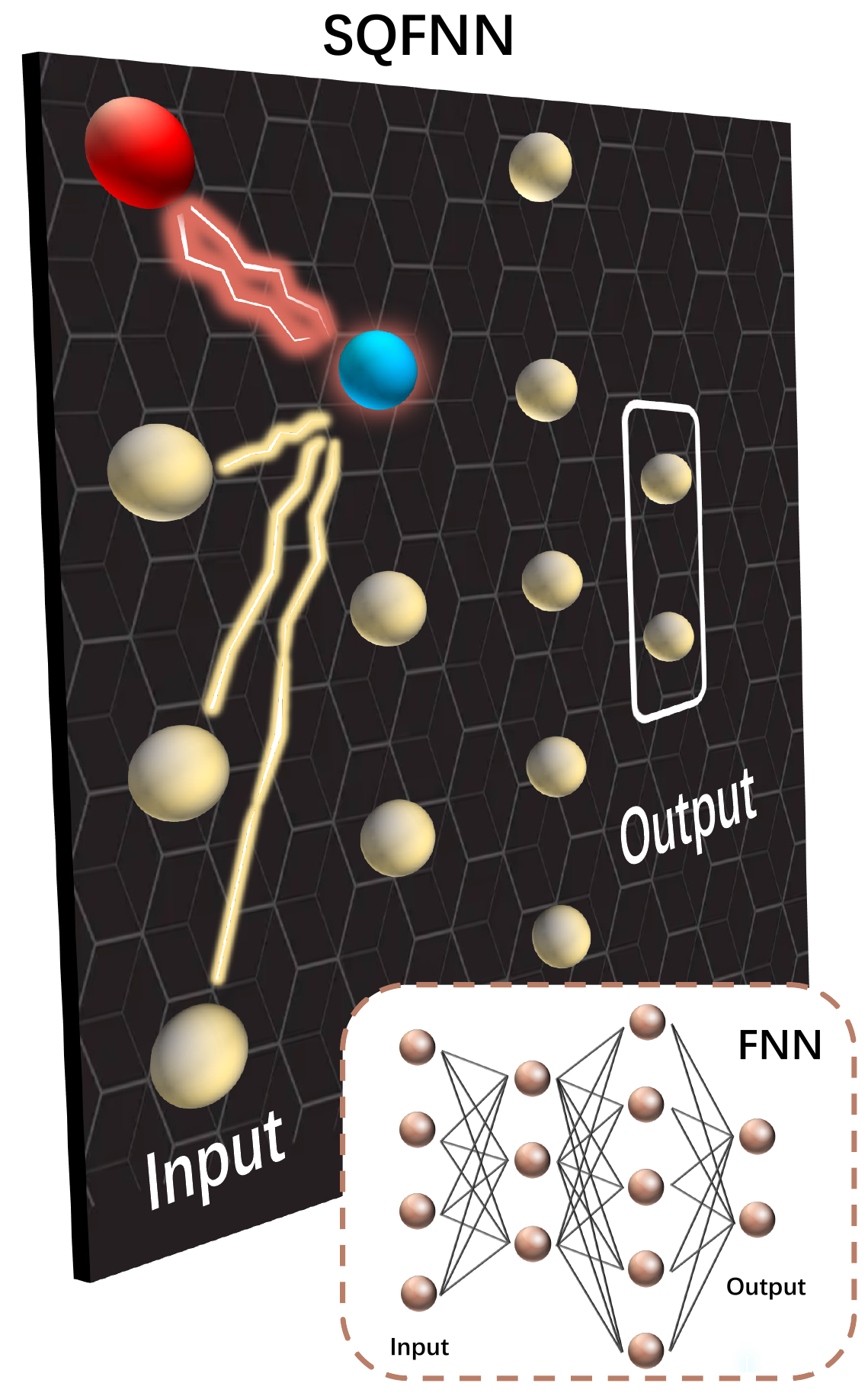}
	\caption{\textbf{The concept of SQFNNs.} The network architecture of a soft quantum feedforward NN (SQFNN) is similar to that of the fully-connected feedforward NN (FNN) displayed at the bottom right corner. There is no feedback in the entire network. Signals propagate unidirectionally from the input layer to the output layer. The first action occurs between the red neuron and the blue neuron. The part in the white box represents the output layer, whose average output is defined as the final output of the network.}
	\label{qnetwork}
\end{figure}

Figure~\ref{qnetwork} shows the concept of an SQFNN. Without loss of generality, we specify that signals propagate from top to bottom and from left to right. Therefore, the evolution equation of the $j$th neuron in the $l$th layer is 
\begin{align}\label{network}%
\rho_{j^{(l)}}^{\text {out}} &\equiv \mathcal{U}_{j^{(l)}}\mathrm{tr}_{\{i^{(l-1)}\} } \notag\\
&(\bigotimes\nolimits_{i^{(l-1)}=1}^{n^{(l-1)}}\mathcal{O}_{i^{(l-1)}j^{(l)}}(\rho_{i^{(l-1)}}^{\text{out}}\otimes\rho^{\text{in}}_{j^{(l)}})),
\end{align} 
where $\mathcal{O}_{i^{(l-1)}j^{(l)}}$ acts on the $i$th neuron in the $(l-1)$th layer and the $j$th neuron in the $l$th layer. The final state of the output layer of the network can be obtained by calculating the final state of each neuron layer by layer with Eq.~(\ref{network}) after considering the local bias superoperator acting upon each neuron. Note that due to the randomness introduced by the measurement operations, the result of a single run of the quantum NN is unstable, i.e., probabilistic. One way to prevent this instability is to obtain the average output of the network by resetting and rerunning the entire network multiple times. This average output is more representative of the prediction made by our quantum NN and is therefore defined as the final output of the network. For each neuron of the output layer, the average output includes the binary outputs in the computation basis and their corresponding probabilities. Although running the network multiple times seems to consume more time and resources, this increase is only equivalent to an additional constant factor on the original consumption \cite{beer2020training} and has no serious consequences. Therefore, running the network multiple times is common for extracting the information of quantum NNs and is also widely adopted by other quantum NN models \cite{beer2020training,farhi2018classification}. Strikingly, this repetitive operation is easy and fast for a quantum computer. For example, the “Sycamore" quantum computer executed an instance of a quantum circuit a million times in $200$ seconds \cite{arute2019quantum}. 

In supervised learning, the NN must output a value close to the label of the training point. The closeness between the output and the label is usually measured by defining a loss function. The loss function in our model can be defined in various ways, e.g., by the fidelity between the output and the expected output or by certain distance measure. In the simulations shown below, a mean squared error (MSE) loss function is adopted, which can be written as
\begin{equation}
\mathcal{L}=\sum_{k=1}^{N} \frac{1}{N}\left|y^k-\tilde{y}^k \right|^2,
\end{equation}
where $N$ represents the size of a training set, $y^k$ represents the label of the $k$-th training point, and $\tilde{y}^k$ represents the predicted label of our network for the $k$-th training point, which is the average value of the output layer of the network obtained by resetting and rerunning the entire network multiple times. This loss function can be driven to a very low value by updating the parameters of the network, thereby improving the network performance. However, the loss function is nonconvex and thus requires iterative, gradient-based optimizers. As information is forwards-propagated in our network, we can use a backpropagation algorithm to update the parameters of the quantum operations. Moreover, since only single-qubit gates are involved in our model, the total number of parameters is not large and is approximately $\sum_{l=1}^{L-1}3\left(n_l+1\right) \times n_{l+1}$, where $L$ is the total number of layers in a network and $n_l$ is the number of neurons in the $l$th layer. This number is directly proportional to the length $L$ of the network and the square of the average width of the network (i.e., the average number of neurons per layer). In particular, the state space involved in computing the gradients is always that of a single neuron, thus circumventing the problem that the state-space size grows exponentially with the number of neurons. Many optimization algorithms widely used in classical NNs are therefore effectively compatible with our quantum NN, such as Adagrad \cite{duchi2011adaptive}, RMSprop \cite{tieleman2012lecture}, and Adam \cite{kingma2014adam}.

Both classical and quantum samples are available for our network, which is similar to other quantum NNs. For classical data, the input features need to be encoded into qubits and fed to the input layer. For quantum data, the quantum states can be decomposed into a tensor product of the qubits in the input layer, as in quantum circuits.

\bigskip
\noindent
\textbf{Simulations} 

In this section, we benchmark soft quantum perceptrons and SQFNNs with simple XOR gate learning, classifying nonlinear datasets and handwritten digit recognition. Our models show extraordinary generalization abilities and robustness to noise in the numerical simulations. \\

\begin{figure*}
	\centering
	\includegraphics[width=16cm]{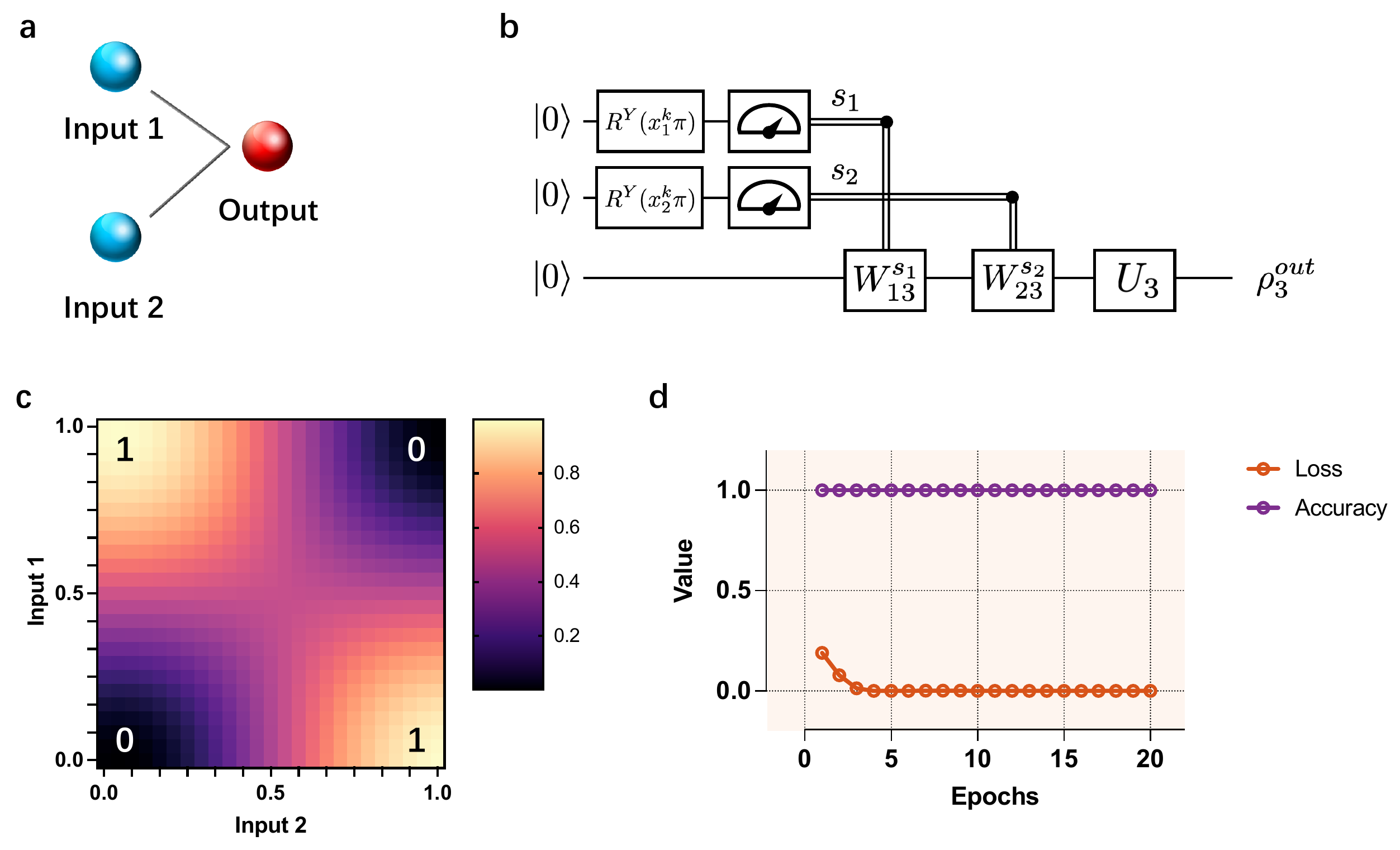}
	\caption{\textbf{Results of XOR gate learning with a soft quantum perceptron. a} The structure of a soft quantum perceptron for learning the XOR gate. The input layer receives two features of the XOR gate, namely, input $1$ and input $2$ of XOR. The output layer predicts the results. \textbf{b} The quantum circuit model corresponding to (\textbf{a}). $R^Y(x_1^k\pi)$ and $R^Y(x_2^k\pi)$ are used to encode the input features (see Methods for details). $W_{13}^{s_1}$, $W_{23}^{s_2}$ and $U_3$ are single-qubit gates with parameters, where the values of the parameters converge during the learning process. \textbf{c} The simulation results of learning the XOR gate. The yellow (black) area represents an output of $1$ ($0$), which is consistent with the truth table of the XOR gate. The true table of the XOR gate is displayed at the four corners of the figure. The soft quantum perceptron fully learns the data structure of the XOR gate. \textbf{d} The training process of the XOR gate. Loss (accuracy) is the value of the loss function (the test accuracy). The soft quantum perceptron achieves $100\%$ test accuracy after the first epoch.} 
	\label{XOR}
\end{figure*}

\begin{figure*}
	\centering
	\includegraphics[width=18 cm]{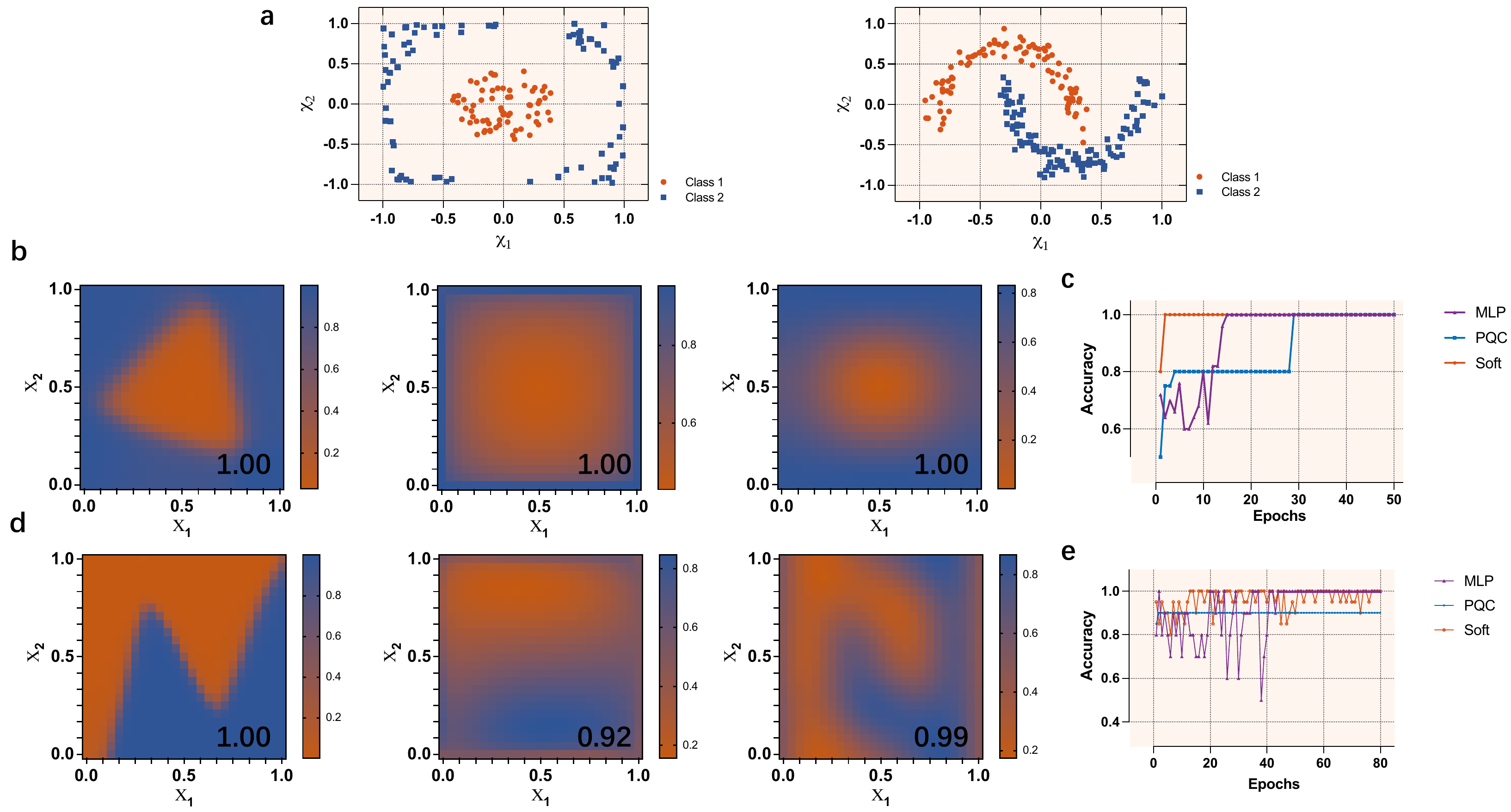}
	\caption{\textbf{Nonlinear decision boundaries by the classical MLP, the PQC model and our model.}
\textbf{a} Displayed from left to right are the visualizations of the ``circles'' dataset and the ``moons'' dataset. $X_1$ ($X_2$) represents the horizontal (vertical) coordinate of the input point. The red (blue) dots represent class $1$ (class $2$). Both datasets are linearly inseparable. \textbf{b} Displayed from left to right are the simulation results for the classical multilayer perceptron (MLP), the PQC model and our model. The classification accuracy is displayed at the bottom right corner of each subfigure. \textbf{c} The training process of learning the ``circles'' dataset with different models. The same results of the ``moons'' dataset are shown in \textbf{d} and \textbf{e}.}
	\label{cir}
\end{figure*}

\begin{figure}
	\includegraphics[width=8.5cm]{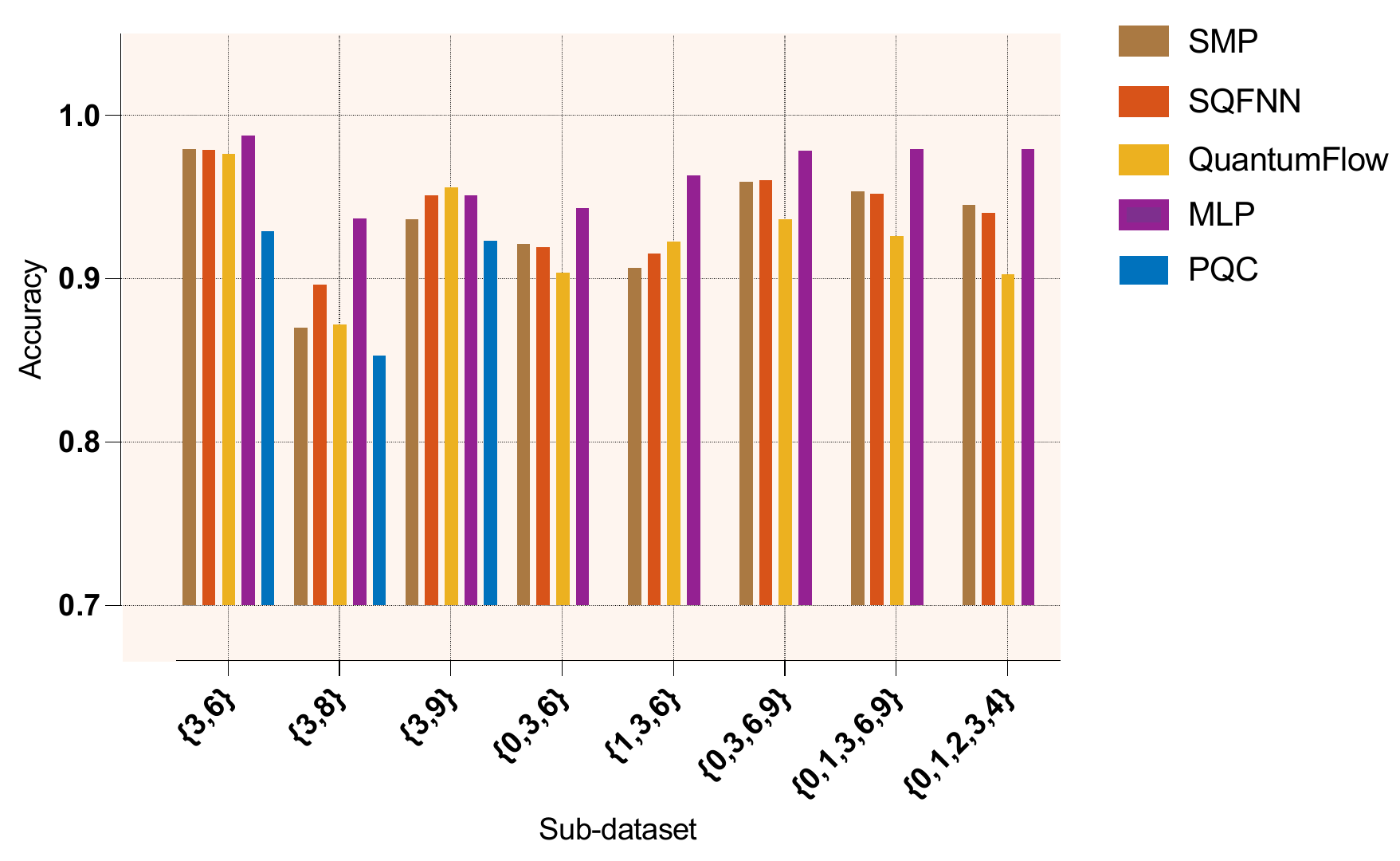}
	\caption{\textbf{Handwriting recognition with the classical MLP, the PQC model, QuantumFlow and our models.} The brown, orange, yellow, purple and blue bars represent the soft multioutput perceptron (SMP), SQFNN, QuantumFlow, the classical MLP and PQC models, respectively. SMP can also be regarded as an SQFNN without hidden layers and accurately identifies handwritten digits that are impossible for classical multioutput perceptrons.}
	\label{hand}
\end{figure}

\noindent
\textbf{XOR gate learning.} The XOR gate is a logic gate that cannot be simulated by classical perceptrons because the input‒output relationship of the gate is nonlinear. Figure~\ref{XOR} reports the results of XOR gate learning with a soft quantum perceptron. Figures~\ref{XOR}\textbf{a},\textbf{b} show the structure and setting of the soft quantum perceptron (see Methods for details). The results clearly show that the soft quantum perceptron is able to learn the data structure of the XOR gate with very high accuracy (Fig.~\ref{XOR}\textbf{c}). Figure~\ref{XOR}\textbf{d} shows the training process, where the training accuracy of our model converges quickly after even one epoch. These results show that our soft quantum perceptron has an extraordinary nonlinear classification ability. 

In addition, we add the bit flip channel, the phase flip channel and the bit-phase flip channel to this task to further demonstrate the performance of our model on realistic quantum systems. We assume that each quantum neuron passes through the same type of quantum noise channel with probability $p$ while waiting to be operated. To make the results more reliable, we repeat the prediction $100$ times with the trained model and use the average accuracy as the evaluation metric. We set the highest noise level in the simulations to $p = 0.5$. Measurements in the simulations are calculated within the limit of the infinite shot number. Details of the simulation results can be found in Table~\ref{XOR_noise} in Methods. The result shows that our model is robust to these different quantum channels. A remarkable result is that our model is fully tolerant to the phase flip channel for the XOR gate learning task. In particular, our model achieves up to $75\%$ accuracy even with a probability of a bit flip or bit-phase flip up to $0.40$. When the probability of a bit flip or bit-phase flip reaches $0.50$, the noise makes qubits $|0\rangle$ and $|1\rangle$ completely indistinguishable. Our model also naturally does not work in this case, which is consistent with theoretical predictions.\\

\noindent
\textbf{Classifying nonlinear datasets.} Two standard two-dimensional datasets (``circles'' and ``moons'') are studied to further demonstrate the ability of soft quantum perceptrons to classify nonlinear datasets and form decision boundaries (see Methods for details). For each dataset, $200$ ($100$) points are generated as the training (test) set. Figure~\ref{cir}\textbf{a} visualizes the training set for the two datasets, where the red (blue) dots represent class $1$ (class $2$). Obviously, the two datasets are linearly inseparable. Figure~\ref{cir}\textbf{b} reports the results of classifying the ``circles'' datasets with different models. Displayed from left to right are the simulation results for the classical multilayer perceptron (MLP), the parameterized quantum circuit (PQC) model and our model. The settings of these models are discussed in detail in the Methods. Figure~\ref{cir}\textbf{c} shows that all three models achieved $100\%$ classification accuracy on the test set of ``circles''. However, the soft quantum perceptron converges faster and learns more robust decision bounds. It is worth reemphasizing that soft quantum perceptrons do not have hidden layers and do not require two-qubit gates.

We also test the tolerance of soft quantum perceptrons for different noise types on this task (see Table~\ref{cir_noise} in Methods). The noise types are added in a manner consistent with the XOR gate learning task described above. The results show that the soft quantum perceptron maintains $100\%$ accuracy on the test set of ``circles'', even when the probability of a bit flip or bit-phase flip is as high as $0.40$. In particular, a soft quantum perceptron can achieve up to $96\%$ accuracy when the probability of a bit flip is as high as $0.49$. In addition, the soft quantum perceptron can maintain over $90\%$ accuracy when the probability of a phase flip is as high as $0.5$. We also found that the robustness of our model can be greatly enhanced when we use SQFNNs. For example, we obtain $100\%$ accuracy when the probability of a phase flip is as high as $0.50$ by adopting a 2-4-2-1 network structure. This suggests that the capabilities of our model can be enhanced by building more complex network structures, which provides strong confidence in handling more complex classification problems with our model.

Figures~\ref{cir}\textbf{d}-\textbf{e} show the results of classifying the ``moons'' datasets with different models. The MLP achieved $100\%$ accuracy, which is slightly higher than the $99\%$ accuracy of the soft quantum perceptron. However, the soft quantum perceptron learns a decision boundary that is better suited to the original data. For comparison, the PQC model can only achieve $92\%$ accuracy. In the experimental setup currently used, our model shows absolute advantages over the PQC model in some tasks.\\

\noindent
\textbf{Handwritten digit recognition.} Finally, we use QuantumFlow, the classical MLP, the PQC model, soft multioutput perceptron (SMP) and the SQFNN to recognize handwritten digits to demonstrate the ability of our models to solve specific practical problems (see Methods for details).
QuantumFlow is a codesign framework of NNs and quantum circuits, and it can be used to design shallow networks that can be implemented on quantum computers~\cite{jiang2021co}. SMP can also be regarded as an SQFNN without hidden layers. The simulation setting is discussed in the Methods section. Figure~\ref{hand} shows the results of different classifiers for classifying different subdatasets from MNIST~\cite{lecun1998gradient}. The results show that the classical MLP performs better than the other four quantum models on all these subdatasets except for $\{3,9\}$. This may be caused by the fact that the classical optimization algorithm has better adaptability to the classical MLP model. Strikingly, the performance of our models (i.e., SMP and the SQFNN) is significantly better than that of QuantumFlow as the number of classes in the dataset increases, implying that our models may have more advantages in dealing with more complex classification problems. For the datasets with two or three classes, our models also perform significantly better than the PQC model and perform comparably to QuantumFlow. For example, the SQFNN achieves $89.67\%$ accuracy on the $\{3,8\}$ dataset, which is $2.47\%$ and $4.34\%$ higher than those of QuantumFlow and the PQC model, respectively. However, our models require only classically controlled single-qubit operations and single-qubit operations, whereas QuantumFlow requires a large number of controlled two-qubit gates or even Toffoli gates to implement the task. In particular, SMP is able to effectively classify handwritten digits with a structure without hidden layers, which is not possible in classical multioutput perceptrons.

\bigskip
\noindent
\textbf{Discussion}

In this work, we develop a new routine for quantum NNs as a platform for quantum neural computing on real-world quantum systems. The proposed soft quantum neurons are subject merely to local or classically controlled single-qubit gates and single-qubit measurements. The simulation results show that soft quantum perceptrons have the ability, beyond that of classical perceptrons, of nonlinear classification. Furthermore, our model is able to classify handwritten digits with extraordinary generalization ability, even in the absence of hidden layers. This performance, combined with the quantum correlations arising from quantum discord in our model, makes it possible to perform nonclassical computations on realistic quantum devices that are extensible to a large scale. Thus, the proposed computing paradigm is not only physically easy to implement, but also predictably exciting beyond classical computing capabilities.

The soft quantum neurons are modelled as independent signal processing units and have more flexibility in the network architecture. Similar to classical perceptrons \cite{goodfellow2016deep,nielsen2015neural}, such units can receive signals from any number of neurons and send their outputs to any number of neurons. This similar property allows our quantum NNs to take classical network architectures that have been proven effective, thereby exploiting the respective advantages of quantum technology and classical network architectures. For example, soft quantum neurons can be combined into quantum convolutional NNs based on convolutional NNs that are widely used in large-scale pattern recognition \cite{krizhevsky2012imagenet}. Moreover, our model enables the construction of quantum-classical hybrid NNs by introducing classical layers. As the final output of our quantum NN is the classical information, part of the classical information can also be processed by classical perceptrons. This advantage makes our model more flexible and thus more adaptable to various problems. 

Our results provide an easier and more realistic route to quantum artificial intelligence. However, some limitations are worth noting. Although the quantum state space involved in computing the gradients in our model is always that of a single neuron, there may also be a barren plateau in the loss function landscape, which hinders the further optimization of the network. Additionally, while soft quantum NNs are much easier to build than standard ones, we need to do more work to understand what kinds of tasks they do well in learning. Future work should therefore include further research on optimization algorithms and building various soft quantum NNs inspired by classical architectures to solve problems that are intractable with classical models.

\bigskip
\noindent\textbf{Methods}

\noindent

\noindent
\textbf{Soft quantum perceptron for XOR gate learning.} We now discuss the details of the simulation setting for XOR gate learning. Figure~\ref{XOR}\textbf{a} shows the model structure for learning an XOR gate, where two neurons in the input layer receive and encode data points, and the neuron in the output layer predicts the outcome. We adopt a simpler and more efficient angle encoding method instead of the method adopted in Ref.~\cite{mitarai2018quantum} to encode the data (Fig.~\ref{XOR}\textbf{b}) and accelerates the convergence of the training process. Specifically, for an input set $\{x^k\}$, we encode the $i$-th feature $x^k_i$ of the $k$-th data point by applying a single-qubit rotation gate $R^Y(x_i^k\pi)$ on the initial qubit $|0\rangle$, where $Y$ represents the rotation along the $Y$ axis and $x_i^k\pi$ represents the rotation angle. Note that a common MSE loss function and the Adam algorithm~\cite{kingma2014adam} are used in the training processes for all tasks in this study. The soft quantum perceptron for learning XOR is optimized for $20$ epochs, and the learning rate is set to $0.1$. Table~\ref{XOR_noise} shows how the test accuracy of our model for the XOR gate learning task varies as the flip probability $p$ increases.\\

\noindent
\textbf{Classifying nonlinear datasets.} A 2-4-1 MLP structure is used for comparison in the task of classifying the ``circles'' dataset, as classical perceptrons are unable to classify nonlinear datasets. The reason for using this structure is that the 2-4-1 MLP needs to learn $17$ parameters, which is approximately the same number of parameters that our model needs to learn. The structure of the PQC model used for comparison is adopted from Ref.~\cite{schuld2020circuit}. This common layered PQC model is denoted as 
\begin{align}\label{PQC}%
U(\bar{\theta})= B_d\left(\bar{\theta}_d\right) \cdots B_{\ell}\left(\bar{\theta}_{\ell}\right) \cdots B_1\left(\bar{\theta}_1\right)
\end{align}
where $\bar{\theta}$ represents the overall learnable parameters of the PQC, $B_{\ell}\left(\bar{\theta}_{\ell}\right)$ is a parameterized block consisting of a certain number of single-qubit gates and entangling controlled gates, and depth $d$ represents the total number of such blocks. These qubits and controlled gates in the same block $B_{\ell}\left(\bar{\theta}_{\ell}\right)$ form a cyclic code. The control proximity range of a cyclic code, denoted as $r$, defines how the controlled gates work. For any qubit index $j \in[0, N-1]$ of an $N$-qubit circuit, the entangling code clock has one controlled gate with the $j$th qubit as the target and the qubit with the index $k = (j+r)\bmod(N)$ as the control qubit (see Ref.~\cite{schuld2020circuit} for details). In each block $B_{\ell}\left(\bar{\theta}_{\ell}\right)$ of our setting, each qubit is acted on by a parameterized universal single-qubit gate. Then, the code block follows. One more optimizable single-qubit gate $R^Y$ acts on each qubit in the final $B_{\ell}\left(\bar{\theta}_{\ell}\right)$. The control proximity range of a cyclic code $r$ is fixed to $1$. Specifically, a $2$-qubit circuit of depth $d=1$ and size $s=6$ is used to classify the ``circles'' dataset, where $d$ is the number of blocks $B_{\ell}\left(\bar{\theta}_{\ell}\right)$ and $s$ is the total number of gates in the circuit other than in the encoding layer. In particular, the encoding method in Ref.~\cite{mitarai2018quantum} is also used in this PQC model for classifying nonlinear datasets. To enrich the expressivity of our model, we adopt the ``parallel encodings'' strategy mentioned in Ref.~\cite{schuld2021effect} when classifying the ``moons'' dataset, that is, using multiple neurons to repeatedly encode the same input in the input layer. In the task of classifying the ``moons'' dataset, we repeatedly encode each input with three neurons. For comparison, we also simulate the results of a 2-10-1 MLP and a $4$-qubit PQC with $d=2$ and $s=24$. The MLP has $41$ parameters to learn. The PQC model also adopts the ``parallel encodings'' strategy in this task. In particular, the soft quantum perceptron does not have hidden layers, so it is a simpler structure compared to the MLP. In fact, in addition to the results presented in the main text, we also found that a $4$-qubit PQC with $d=10$ and $s=40$ could only achieve $94\%$ accuracy when classifying the ``moons'' dataset. Table~\ref{cir_noise} shows how the test accuracy of our model for classifying the ``circles'' datasets varies as the flip probability $p$ increases. Note that this effect is continuous, but the presentation of our results in a discrete table format may create an impression of discontinuity. Moreover, when the probability of bit flip or bit-phase flip reaches $0.5$, the $|0\rangle$ and $|1\rangle$ components of the corresponding quantum state become indistinguishable in the computational basis, resulting in the inability to extract any relevant information. This causes a sudden drop in the probability of successful learning. This effect is particularly pronounced in close proximity to the $0.5$ probability threshold.\\

\begin{table}
\center
\begin{tabular}
{c @{\hspace{0.35cm}} c @{\hspace{0.35cm}} c  @{\hspace{0.35cm}} c @{\hspace{0.35cm}} c @{\hspace{0.35cm}} c @{\hspace{0.35cm}} c }  \hline \hline 
 \multirow{2}*{Noise channels}&\multicolumn{6}{c}{Flip probability}\\    
&$0.10$  & $0.20$& $0.30$&  $0.35$& $0.40$ & $0.50$\\
 \hline
 Bit flip & $100\%$ & $100\%$ & $100\%$ & $100\%$ & $75\%$ &$50\%$ \\
Phase flip & $100\%$ & $100\%$ & $100\%$ & $100\%$ & $100\%$ &$100\%$ \\
Bit-phase flip & $100\%$ & $100\%$ & $100\%$ & $100\%$ & $75\%$ &$50\%$ \\
 \hline \hline
\end{tabular}
\caption{Test accuracies of learning the XOR gate with the soft quantum perceptron after a bit flip channel, phase flip channel, or bit-phase flip channel with different flip probabilities.}
\label{XOR_noise}
\end{table}

\begin{table}
\center
\begin{tabular}
{c @{\hspace{0.35cm}} c @{\hspace{0.35cm}} c  @{\hspace{0.35cm}} c @{\hspace{0.35cm}} c @{\hspace{0.35cm}} c }  \hline \hline 
 \multirow{2}*{Noise channels}&\multicolumn{5}{c}{Flip probability}\\
&$0.10$  & $0.20$& $0.30$& $0.40$ & $0.50$\\
 \hline
 Bit flip & $100\%$ & $100\%$ & $100\%$ & $100\%$ & $32\%$ \\
Phase flip & $100\%$ & $100\%$ & $93\%$ & $92\%$ &$91\%$ \\
Bit-phase flip & $100\%$ & $100\%$ & $100\%$ & $100\%$ & $68\%$ \\
 \hline \hline
\end{tabular}
\caption{Test accuracies of learning the ``circles'' dataset with the soft quantum perceptron after a bit flip channel, phase flip channel, or bit-phase flip channel with different flip probabilities.}
\label{cir_noise}
\end{table}

\noindent
\textbf{Simulation setting for handwritten digit recognition.} The specific simulation seting is as follows. First, we extract several subdatasets from MNIST. For example, $\{3,6\}$ represents the subdataset containing two classes of the digits $3$ and $6$. After that, we apply the same downsampling size to all images from the same subdataset of MNIST. Specifically, we downsample the resolution of the original images from $28 \times 28$ to $4 \times 4$ for the datasets with two or three classes, and to $8 \times 8$ for datasets with four or five classes. Finally, we use the structure from Ref.~\cite{jiang2021co} that contains a hidden layer for QuantumFlow, the classical MLP, and the SQFNN, where the hidden layer contains $4$ neurons for two-class datasets, $8$ neurons for three-class datasets, and $16$ neurons for four- and five-class datasets. The input and output layers of these models (including SMP) are determined by the downsampling size and the number of digits in the subdatasets. Note that the PQC model is designed as a $4$-qubit circuit of $d=10$ and $s=120$ due to the lack of the concept of neurons. The PQC model is usually used as a binary classifier in the current study. Therefore, the PQC model is only used to classify the datasets with two classes in this task. Other QuantumFlow settings, such as accuracy, are consistent with those in Ref.~\cite{jiang2021co}.

\bigskip
\noindent\textbf{Data Availability}\\
Data generated and analyzed during the current study are available from the corresponding author upon reasonable request.\\

\bigskip
\noindent\textbf{Conflicts of Interest}\\
The authors declare that there is no conflict of interest regarding the publication of this article.

\bigskip
\noindent\textbf{Authors' Contributions}\\
Z.-B.C. conceived and supervised the study. M.-G.Z., Z.-P.L, H.-L.Y., and Z.-B.C built the theoretical model. M.-G.Z., Z.-P.L, and H.-L.Y. performed the simulations. M.-G.Z., Z.-P.L, H.-L.Y., and Z.-B.C cowrote the manuscript, with inputs from the other authors. All authors have discussed the results and proofread the manuscript.\\

\bigskip
\noindent\textbf{Acknowledgements}\\
We gratefully acknowledge the supports from the National Natural Science Foundation of China (No.
12274223), the Natural Science Foundation of Jiangsu Province (No. BK20211145), the Fundamental Research Funds for the Central Universities (No. 020414380182), the Key Research and Development Program of Nanjing Jiangbei New Aera (No. ZDYD20210101), and the Program for Innovative Talents and Entrepreneurs in Jiangsu (No. JSSCRC2021484).\\


\end{document}